# MODELLING OF DESORPTION FROM SMALL METAL CLUSTERS

M. KOLEVA*, L. PETROV

*Institute of Catalysis, Bulgarian Academy of Sciences, 1113 Sofia, Bulgaria*
*Fax: (+359 2) 756116, e-mail: mkoleva@.ic.bas.bg*

ABSTRACT

It is found out that there are physical effects, typical of small metal clusters deposited on a non-adsorbing support, that cause a difference between the adsorption properties of sites at the metal cluster surface and those of sites at the boundary of metal-support. This difference gives rise to an extra band in the IR spectra of the adsorbate whose major property is that its intensity changes with cluster size.

The migration of chemisorbed molecules (atoms, ions) and molecules (atoms, ions) in any excited state over adsorption sites of different type results in changes in the state of the molecule (atom, ion) since the latter adjusts its state to the set of levels of another type of site. This can be described figuratively as a transformation of the type of adsorption site. Thus, a single type of adsorption site, called next surface effective site (SES), is formed. The energy of desorption from SES site retains an explicit dependence on the metal cluster size. A simulation of TPD spectra of oxygen desorption from small Pt clusters deposited on NaX zeolite was carried out. A comparison to experimental TPD spectra obtained by Jaeger and co-workers for the same system is made.

*Keywords*: clusters, desorption, TPD spectra, platinum, oxygen, zeolites.

## 1. INTRODUCTION

So far, the theoretical investigations of the macroscopic adsorption characteristics, such as sticking coefficient and desorption rate, have been developed for systems of infinite size. It is to be expected, however, that the adsorption properties of small metal clusters, deposited on a non-adsorbing support, differ from the adsorption properties of infinite well-defined single crystal surfaces. A specific property of the small clusters is related to effects that act toward formation of difference between the properties at the cluster surface and at the boundary of metal-support. These effects, which are typical of a large class of systems, like adsorbate/metal cluster/support, are the following. (i) The screening effect of the electrostatic field of the support is larger for adsorption sites at the cluster surface than for adsorption sites at the metal-support boundary. (ii) The presence of the boundary acts toward an increase of the electron density localization, an analogue to the Schottky effect.

The difference between the adsorption properties at the surface of the cluster and at the boundary of metal-support becomes evident when the adsorption properties at the cluster surface are insensitive to the particular local geometry of the surface and the major adsorption mechanism is the electronic transfer. Then, only two types of adsorption sites are formed, namely, adsorption sites at the cluster surface and adsorption sites at the boundary of the support. On the contrary, when the adsorption properties are sensitive to the local geometry, formation of several types of adsorption sites is possible. Thus, the effect of the difference between the surface and the boundary types of adsorption sites is blurred.

An example of a system that meets the above confinements is oxygen adsorbed on Pt or Pd clusters supported on zeolite lattices. According to Vedrine et al. [1], the major mechanism of interaction between a Pt cluster and NaY zeolite is the electronic transfer. It is well established that the zeolites form an intense electrostatic field. More precisely, Vedrine et al. [1] have found that: (i) Pt atoms are ionic-bonded to the lattice oxygen; (ii) no difference in electronic properties is observed for 1 nm and 2 nm Pt clusters in the zeolite; (iii) Pt clusters strongly interact with the zeolite by electronic transfer; (iv) Pt and Pd ions at cationic sites on the zeolite are shown by XPS to be bonded to lattice oxygen by ionic bonds while more covalent bonds are involved in the case of Pt or $Pt(NH_3)_4^{2+}$ complexes and to a large extent in the case of PtO and PdO; (v) electron acceptor sites develop on NaY as Pt content increases.

Kohiki and Ikeda's [2] evaluation shows that the influence of the substrate on the electron core binding energy at small Pd clusters is relatively large.

Item (iv) confirms the assumption about extra-localization of the electron density at the boundary of cluster-support. It is well established that the electron density localization is a premise for covalent bond formation. That is why it is supposed that PtO is formed at the boundary as follows: a Pt atom at the boundary, bonded to lattice oxygen, is bound to another oxygen atom that belongs to the adsorbate. Therefore, one should expect the appearance of an extra band in the IR spectra of adsorbed oxygen. The position of the band corresponds to the bond energy of $PtO_2$ while its intensity changes with cluster size. The dependence of the intensity on cluster size is a result of the assumption that PtO content is proportional to the number of occupied boundary adsorption sites but not to the whole number of occupied adsorption sites.

The formation of two major types of adsorption sites is supported also by calorimetric investigations. Norman et al. [3] have found that the energy of PtO is $355 \pm 41.8$ kJ/mol whereas Savitskii et al. [4] have reported that the minimum heat of adsorption of oxygen on Pt clusters is 125 kJ/mol.

It is experimentally well established that the sticking coefficient of oxygen adsorbed on different well-defined surfaces is insensitive to the crystallographic orientation. This fact supports the assumption that only two types of adsorption sites are formed at the Pt clusters deposited on NaX zeolite.

The principal goal of the present paper is to study the rate of oxygen desorption from small Pt clusters deposited on NaX zeolite. The particular interest in the desorption rate is provoked by the unusual properties of the TPD spectra obtained by Jaeger et al. [5,6] and by the above-listed strong experimental evidence of formation of only two types of adsorption sites. An important feature of the experimental data is that each TPD spectrum contains a single peak whose position changes with cluster size. The assumption that the desorption from both types of adsorption sites are two independent processes would has resulted in TPD spectra with two well resolved peaks whose relative intensity would have changed with cluster size but their position would have been fixed. The independence of desorption from different types of adsorption sites is possible when desorption is completed in the period of a single vibration as this is supposed on deriving the Wigner-Polyani equation for the desorption rate. This assumption, however, is too tight to be relevant to every adsorption system. Next, it is assumed that desorption is a process of gradual excitation from ground (chemisorbed) state to a free particle (desorbed state). The dominant property is that the rate of surface migration at each level of excitation is non-vanishing. Therefore, an excited molecule (atom, ion) can visit several adsorption sites of different type during desorption. Occurring on another type of site, the excited molecule (atom, ion) adjusts its state to the nearest state that belongs to the set of levels of the new site. Next, the process of adjustment is figuratively called transformation of the type of adsorption site. Apparently, this process affects the overall energy of desorption. The possibility of undergoing transformation breaks the statistical independence of both types of adsorption sites and a single adsorption site, called surface effective site (SES), is figuratively formed. An explicit equation for desorption rate from SES's is discussed in the next section. Further, a comparison between simulated TPD spectra, according to the derived equation, and TPD spectra experimentally obtained by Jaeger et al. [5,6] is made.

## 2. BACKGROUND FOR DERIVATION OF THE DESORPTION RATE EQUATION

The main hypothesis of the present paper is that desorption is a process of gradual excitation of a chemisorbed molecule (atom, ion) from the ground state to a state of a free molecule (atom). The process includes excitation of the electronic, vibration and rotational degrees of freedom of a chemisorbed molecule (atom, ion). Desorption can be described as a gradual excitation through a sequence of levels that are formed by the Hamiltonian of the interaction between a chemisorbed molecule (atom, ion) and an adsorption site. A distinctive common property of chemisorption is that there are four open channels at any level of excitation. The molecule (atom, ion) can be excited to a higher level, it can migrate to a neighbouring adsorption site, it can stay at the same site at the same level or it can relax to a lower level. Since the chemisorption proceeds on metal clusters deposited on a solid support, the principal mechanism of migration both in the ground and in any excited state is the phonon-assisted hopping. Usually, the rate of desorption is considerable at high enough temperatures such that the hopping diffusion coefficient is also relatively large. The latter provides the importance of the following effects:

(i) Each type of adsorption site retains a specific set of excited levels. On hopping to a site of another type, a molecule (atom, ion) changes its state by excitement or relaxation to the nearest level that belongs to the set of levels of the latter type of adsorption site. As it was mentioned above, this process is figuratively called a transformation of the type of adsorption site. Apparently, the number of transformations that a molecule undergoes during desorption affects the overall energy of desorption. This problem is discussed in the next subsection.

(ii) When the adsorbate forms submonolayer coverage, another type of transition is possible. While a molecule (atom, ion) is in an excited state on an unoccupied adsorption site, a chemisorbed molecule (atom, ion) from a neighbouring site can hop to the same site. Then the excited molecule (atom, ion) cannot anymore relax to a lower state (and eventually being chemisorbed on that site) but it can only either migrate or be excited to a higher level. The occupation of the adsorption site by the chemisorbed molecule (atom, ion) results in a qualitative change of the Hamiltonian of the interaction between the excited molecule (atom, ion) and the unoccupied adsorption site. Therefore, two different sets of excited states are formed: one is formed when the excitation occurs on an unoccupied adsorption site and the other one is formed on an occupied adsorption site. Further, the former set is called a set of intrinsic excited levels while the latter one is a set of extrinsic excited levels. Transitions between them are possible due to the migration of the chemisorbed molecules. The influence of these transitions on desorption rate is discussed in the second subsection.

### 2.1 SURFACE EFFECTIVE SITE

The transformation of the type of adsorption site is possible because the system of excited levels is specific for each type of adsorption site. During migration an excited molecule (atom, ion) changes its state by excite-ment or relaxation to the nearest level that belongs to the set of levels of the other type of adsorption site. Therefore, any chemisorbed molecule (atom, ion) can start desorption from adsorption site of a given type and complete it on a site of another type by undergoing several transformations. The actual energy of desorption depends on the overall probability of visiting sites of each type. The state of occupation of a given adsorption site does not affect this probability because the set of levels is specific for each type of adsorption site both for intrinsic and extrinsic states.

The main goal of the present paper is to evaluate the rate of oxygen desorption from small Pt clusters deposited on NaX zeolite. The reason for this particular choice is that this system meets all the conditions listed in the Introduction. Therefore, two types of adsorption sites are formed: adsorption sites at the Pt cluster surface and adsorption sites at the metal-support boundary. As it was mentioned above, the sticking coefficient of oxygen on well-defined infinite surfaces is almost independent of the crystallographic orientation. This allows an assumption of approximating the Pt clusters to hemispheres of size L and of random distribution of identical adsorption sites on Pt cluster surfaces. It is also assumed that the boundary adsorption sites retain identical properties and that they are randomly distributed on the boundary.

It is supposed that a recombination of the chemisorbed oxygen atoms occurs when two atoms are in the ground (or low excited) state. That is why the subject of the considerations is the further behaviour of an excited oxygen molecule.

The number of transformations of the type of adsorption site depends on the probability of visiting any adsorption site by migration. When the following estimation holds any excited molecule visits any adsorption site with equal probability:

$$\frac{L^2}{a^2}\tau_D \ll \tau_{des},\qquad(1)$$

where L is the size of the cluster, $a$ is the typical nearest neighbour distance, $\tau_D$ is the time of a single hop at an excited state, and $\tau_{des}$ is the desorption duration. The slowest process of excitation is the excitation of the vibration degrees of freedom by excitation of the rotational degrees of freedom. According to Zhdanov [7], the duration of this process for diatomic molecules is of the order of $10^{-6}$ s. The highest typical value for the time of single hop of chemisorbed oxygen on Pt surface is of the same order. However, the migration rate in an excited state, $\tau_D$, is of several orders higher magnitude than that in the ground (chemisorbed) state and since the size of the clusters considered is 1–2 nm, it is evident that Eq. 1 holds. Since any adsorption site is visited with equal probability, the overall probability of desorption from each type of adsorption site is proportional to the number of sites of a given type.

The energy of desorption from the SES site comprises the energies of desorption from the two types of sites but they enter with corresponding statistical weights. The latter are defined by the probability of undergoing a transformation. Under the condition of Eq. 1, the statistical weights are proportional to the relative number of sites of each type. The energy of desorption from SES due to this type of transformation reads:

$$E_d^{(1)} = \mu E_{ds} + (1-\mu)E_{db} \qquad(2)$$

where $\mu$ is the relative number of the surface type adsorption sites; $E_{ds}$ is the energy of desorption from the surface type adsorption sites; $E_{db}$ is the energy of desorption from the boundary type sites. Under the approximation of Pt clusters to hemispheres $\mu$ is defined by:

$$\mu = \frac{N_s}{N_s + N_b} \approx \frac{1}{1+\frac{a}{L}}, \qquad(3)$$

where $N_s$ is the number of the surface type-sites; $N_b$ is the number of the boundary type-sites.

Another type of adsorption site transformation is possible when $E_{db} > E_{ds}$. These transformations occur when a chemisorbed molecule on the surface reaches an occupied boundary site. More precisely, when a chemisorbed molecule on the surface reaches a boundary site by migration, the following alternatives are possible:

(i) The boundary site is occupied and the molecule on this site is in the ground state. Then the molecule that comes from the surface is scattered back;

(ii) The boundary site is unoccupied. Then the molecule that comes from the surface occurs in an excited state. This state, however, is a low excited one since the molecule migrates on the surface, being at the ground state. Therefore, the probability of relaxation to the ground state is much higher than the probability of desorption, so the surface molecule is most probably readsorbed on an unoccupied boundary site.

(iii) If a molecule that comes from the surface reaches a boundary site on which the molecule is in highly excited state, an inelastic collision between both molecules takes place. The collision results in desorption of one of the molecules. The molecule that comes from the surface is in a low excited state while the other molecule has some vibration and rotational states excited as well as some electronic states excited. Thus, most likely the molecule from the surface becomes in an intrinsic state with the energy of desorption $E_{db} - E_{ds}$ while the boundary molecule is scattered to the state of a free molecule, i.e. it is desorbed.

The probability per unit time (frequency) of this transformation to take place equals the frequency of a chemisorbed molecule on the surface to reach the boundary by starting anywhere on the surface times the frequency of the boundary molecule to be at a highly excited intrinsic state. The frequency of reaching the boundary by starting anywhere on the surface is:

$$\frac{a}{L} \qquad (4)$$

Next, it is found that for the system considered the following estimation holds:

$$\frac{L}{a}\tau_D^* \ll \tau_{des}^b, \qquad (5)$$

where $\tau_D^*$ is the single hop duration of the surface molecule between two neighbouring boundary sites. Since the surface molecule is in an excited state at the boundary, the duration of a single hop is much shorter than the duration of a single hop of migration at the ground state. $\tau_{des}^b$ is the duration of desorption at a boundary site. Considerations for the values of $\tau_D^*$ and $\tau_{des}^b$ are given above (see Eq. 1). It is obvious that for oxygen desorption from Pt clusters of 1–2 nm size estimation (5) holds. The fulfilment of Eq. 5 ensures that a surface molecule can reach any boundary site before being desorbed. Formally, this provides an independence of the probability of these transformations on coverage. Therefore, the frequency of the transformation to occur reads:

$$\frac{a}{L}\nu_0 \exp\left(-\frac{E_{db}}{RT}\right), \qquad (6)$$

where $\nu_0$ is the frequency factor.

The contribution of this type of transformation to the energy of the SES site is:

$$E_d^{(2)} = \frac{a}{L}\nu_0 \exp\left(-\frac{E_{db}}{RT}\right)(E_{db} - E_{ds}). \qquad (7)$$

It is worth noting that both types of transformations are independent. The former type takes place because any molecule in any excited state can change the type of the site. The latter type occurs when a molecule chemisorbed on the surface reaches the boundary by migration in the ground state. The specific collision described above takes place and results in desorption of already excited molecule from a boundary site. Thus, the latter type of transformation is typical of specific spatial distribution of the types of adsorption sites at small clusters for which $E_{db} > E_{ds}$ holds.

The lateral interactions of the chemisorbed molecules can only modify $E_{db}$ and $E_{ds}$ but they do not induce another type of transformation. That is why they are not considered in the present paper.

The overall energy of desorption from a surface effective site becomes:

$$E_d = \mu E_{ds} + (1-\mu)E_{db} + \frac{a}{L}\nu_0 \exp\left(-\frac{E_{db}}{RT}\right)(E_{db} - E_{ds}). \qquad (8)$$

It is obvious that when $L \to \infty$ $E_d \to E_{ds}$ which is a result of the specific spatial distribution of the types of adsorption sites on the Pt clusters. When L is finite, there is an apparent dependence of $E_d$ on the cluster size, L.

## 2.2. INTRINSIC-EXTRINSIC TRANSITIONS

Transitions of intrinsic state to extrinsic state occur when two molecules, one of which is in the ground state and the other one is in an excited state, occur on one and the same adsorption site after migration. Since no more than one molecule can be at the ground state, i.e. in a chemisorbed state, the molecule in the excited state can only migrate or be excited to a higher state but it cannot relax (and eventually be chemisorbed) on that adsorption site. The extrinsic states are more highly excited states than the intrinsic ones. That is why these transitions increase the probability of desorption. The rate of migration of the chemisorbed molecules increases when the temperature is raised, which ensures an enhanced rate of these transitions in the desorption process.

The specific spatial distribution of both types of adsorption sites at the Pt clusters and the suggestion $E_{db} > E_{ds}$ provide that only transitions that occur on the cluster surface do substantially contribute to the desorption rate. The major mechanism of chemisorbed molecule diffusion is the phonon-assisted hopping. The energy of activation of the hopping is proportional to the energy of the ground state. Since $E_{db} > E_{ds}$, the hopping rate of the molecules chemisorbed at the boundary is much smaller than the hopping rate of the molecules chemisorbed at the surface. That is why only transitions that take place at the surface are considered next.

The systems of levels of the intrinsic and the extrinsic states are qualitatively different because they have different number of open channels. Therefore, it is next assumed that only two levels of qualitatively different properties can approximate both systems, namely:

(i) intrinsic level that retains three open channels: migration (probability $f_m$), further excitation (probability $f_d$) and further relaxation (probability $f_a$).

(ii) extrinsic level that retains two open channels: migration (probability $f_m^*$) and further excitation (probability $f_d^*$).

The probabilities of different channels at both levels are parameters in the following considerations. They can be carried out by any quantum-mechanical approach available to the system considered.

First King [8] introduced this approximation to evaluate the influence of the excited states mobility on the sticking coefficient and the desorption rate. He assumed that an excited molecule makes an infinite number of hops while being desorbed. This approximation is justified by the rapid conversion of the pre-exponential factor of King's equation for the desorption rate:

$$F_{\inf}^*(\theta) = f_d + f_m\left[(1-g(\theta))f_d^* + g(\theta)f_d\right] + \\ f_m\left[(1-g(\theta))f_d^* + g(\theta)f_d\right]\left[(1-g(\theta))f_m^* + g(\theta)f_m\right], \qquad (9)$$

where $\theta$ is the coverage; $g(\theta)$ is the probability of an adsorption site to be available for adsorption. According to King $g(\theta) = (1-\theta)^m$, where $m$ is the order of associa-tion). Thus, King takes into account only whether an adsorption site is occupied or not but he does not involve intrinsic-extrinsic transitions. A modification of King's expression is made which is twofold: (i) any migration of a cluster surface molecule can make only a finite number of hops. The number of possible hops is restricted since, reaching the boundary, the molecule is either readsorbed or scattered to a state of a free molecule (desorbed state); (ii) the impact of the intrinsic-extrinsic transitions on $g(\theta)$ is to be carried out. The intrinsic-extrinsic transitions cause a decrease of $g(\theta)$ since an unoccupied adsorption site can become occupied and an excited molecule on this site cannot be further adsorbed on the same site.

First, a restriction over possible number of hops is taken into account. The average number of hops of a molecule excited anywhere on the cluster surface is $N = L^2/a^2$. Therefore:

$$F(\theta) = F_{\inf}^*(\theta) - f(\theta), \qquad (10)$$

where the separation into two parts is formal. After some rearrangement of Eq. 9:

$$F_{\inf}^*(\theta) = f_m \left\{ \sigma + KS(\theta) \frac{1 - \Gamma g(\theta)}{S(0) g(\theta)} \right\} \qquad (11)$$

$$f(\theta) = \frac{S(\theta)}{S(0)(1 - \Gamma g(\theta))} \left( \frac{f_d^*}{K} \right)^N (K - g(\theta)), \qquad (12)$$

where:

$$\sigma = \frac{f_d}{f_m}, \qquad K = \frac{f_d^*}{f_m^* - f_m}, \qquad \Gamma = f_d^* - f_d. \qquad (13)$$

$S(\theta)$ is the sticking coefficient in the frame of the proposed model:

$$S(\theta) = \frac{\sigma f_a}{(f_m^* - f_m)} \frac{1}{(1 + Kg(\theta))}. \qquad (14)$$

$\Gamma$, $\sigma$ and $K$ vary restrictively by obvious constraints: $f_d^* + f_m^* = 1$ and $f_a + f_m + f_d = 1$. Since the extrinsic state is much more weakly bonded than the intrinsic one, it is assumed that $f_d^* \gg f_d$ and $f_m^* \gg f_m$. Then:

$$\sigma \ll 1, \quad \Gamma \approx 1, \quad K = \frac{1 + \Gamma - \sigma f_m - f_a}{1 - \Gamma}. \qquad (15)$$

It is obvious that with certain parameter values, $f(\theta)$ is of the order of $F_{\inf}^*(\theta)$ and thus it cannot be neglected.

The next item is modifying $g(\theta)$ by involving intrinsic-extrinsic transitions. The qualitative change of the Hamiltonian that is caused by this transition gives a difference in energy between the intrinsic and the extrinsic state. Therefore, the modified $g(\theta)$ becomes:

$$g(\theta) = (1 - \theta)^m \exp\left( -\frac{z(\theta)\varepsilon}{RT} \right), \qquad (16)$$

where $\varepsilon$ is equal to the energy difference between the intrinsic and the extrinsic state, $z(\theta)$ is equal to the probability that an intrinsic-extrinsic state transition occurs on a single site, $T$ is the temperature of the cluster surface. The derivation of $z(\theta)$ has been made in Refs. 9 and 10. The basic factor that determines $z(\theta)$ is the value of the $\tau_p/\tau_{d0}$ ratio, where $\tau_p$ is the lifetime of the intrinsic state and $\tau_{d0}$ is the duration of a single hop of a chemisorbed molecule. Provided the chemisorbed molecules are randomly distributed over the adsorption sites, the following extremes are possible.

(i) If $\dfrac{\tau_p}{\tau_{d0}} \approx 1$, then

$$z(\theta) = (2z_0 - 2) \exp\left( -\frac{\theta_s}{\theta} + 1 \right), \qquad (17)$$

where $z_0$ is the average number of nearest neighbours to a single adsorption site, $\theta_s$ is the coverage at saturation.

(ii) If $\dfrac{\tau_p}{\tau_{d0}} \ll 1$, then

$$z(\theta) = (2z_0 - 2) \sum_{n=1}^{\infty} \left( \frac{\theta}{\theta_s} \right) \exp\left( -\frac{\theta_s}{\theta} \right). \qquad (18)$$

(iii) If $\dfrac{\tau_p}{\tau_{d0}} \gg 1$, then $g(\theta)$ is derived by the quasi-chemical approach.

The rate of association $m$ is 2 for all the extremes.
Finally, the desorption rate reads:

$$r_d = -v_0 F(\theta) \exp\left( -\frac{E_d}{RT} \right), \qquad (19)$$

where $F(\theta)$ and $E_d$ are given by Eqs. 10 and 8, respectively. The most distinctive property of the desorption rate, presented by Eq. 19, is the apparent dependence of the desorption energy on cluster size.

## 3. SIMULATION OF TPD SPECTRA AND COMPARISON WITH THE EXPERIMENT

The aim of this section is to stimulate TPD spectra of oxygen desorption from Pt clusters deposited on NaX zeolite. The choice of this particular system is associated with the following features of the experimentally obtained TPD spectra: (i) the TPD peaks shift toward lower temperatures on increasing the average size of Pt clusters (Fig. 1); (ii) the position of the peaks slightly shifts toward higher temperatures on increasing the heating rate (Fig. 2); (iii) the position of the peaks does not shift on changing the initial coverage (Fig. 3).

Jaeger et al. [5,6] have not succeeded in explaining these features under the suggestion that desorption proceeds through several independent types of adsorption sites at reasonable values of the fitting parameters.

The desorption rate presented by Eq. 19 is relevant for the system considered when oxygen readsorption in the zeolite cavities is negligible. The readsorption is to be expected when there are macroscopic areas of equal pressure of outgoing gases in the zeolite lattice. The realization of such case is hardly to be expected for ordered structure, such as the zeolite lattice, under conditions specific of experiment: the rate of outgoing flow is relatively low, so the flow can be assumed laminar.

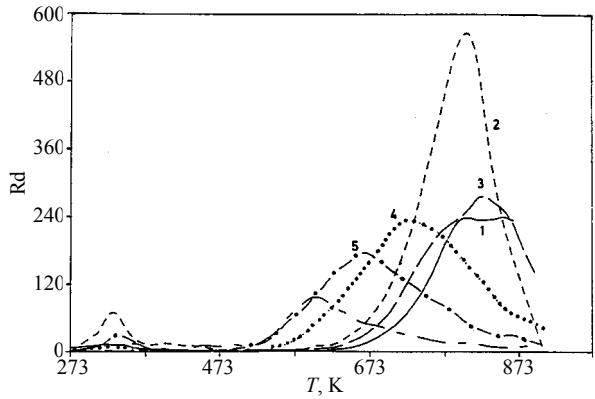

**Fig. 1.** Experimental TPD spectra for samples with the following average Pt cluster size, L, and initial coverage, $\Theta$: curve (1) L=1.1 nm, $\Theta$=0.138; curve (2) L=1.6 nm, $\Theta$=0.216; curve (3) L=1.6 nm, $\Theta$=0.346; curve (4) L=2.8 nm, $\Theta$=0.246; curve (5) L=4.2 nm, $\Theta$=0.272; curve (6) L=5.0 nm, $\Theta$=0.116 (reproduced with permission from Ref. 5)

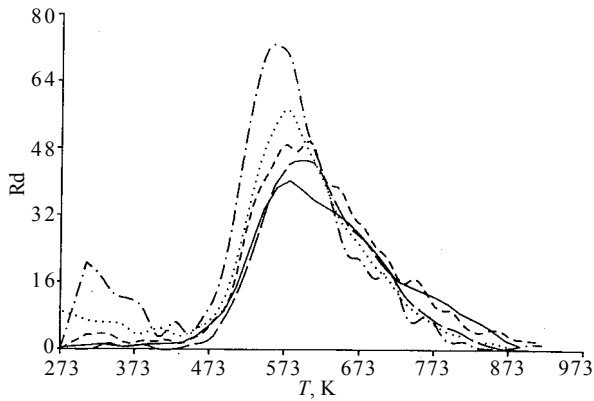

**Fig. 2.** Experimental dependence of the TPD spectra on the heating rate with sample 6; heating rate (K/min): (——) 57.5; (– – –) 23.0; (— —) 11.5; (. . .) 5.75; (—.—.—) 1.15 (reproduced with permission from Ref. 6)

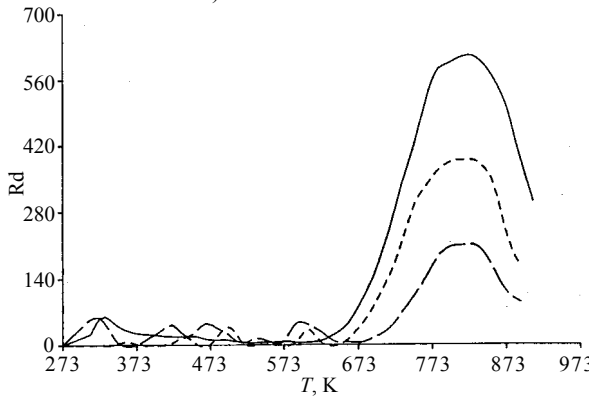

**Fig. 3.** Experimental dependence of the TPD spectra on the initial coverage with sample 3, $\Theta$ %; (——) 34.6; (– – –) 19.8; (— —) 9.2; (reproduced with permission from Ref. 6)

An issue of primary importance is that the best fit between simulated and experimental TPD spectra is to use reasonable values of the fitting parameters. To diminish the number of the fitting parameters the desorption energies of both types of adsorption sites are fixed, namely 355 and 125 kJ/mol for the boundary adsorption sites and the surface type adsorption sites, correspondingly. The average distance between neighbouring adsorption sites is fixed at 0.3 nm, which well approximates the experimentally obtained 0.277 nm [11]. Therefore, the fitting parameters are $v_0, \varepsilon, \Gamma, \sigma$ and $K$. $v_0$ is defined by the edge of the phonon spectrum, whose value for small clusters is about an order of magnitude less than the value of the edge of infinite system [12]. The best fit is achieved at $v_0 = 8 \times 10^{12} s^{-1}$, which satisfies this confinement. The values of the best fit of the other parameters are:

- $\Gamma = 0.999$; $\sigma = 0.005$; $K = 12$. These values satisfy Eq. 15;

- $\varepsilon = 0.9$ kJ/mol. The same value is achieved for the systems $O_2$/Pt(110) [9] and $O_2$/Si(100) [10]. The uniformity of $\varepsilon$ for these different systems supports the suggestion that the intrinsic and extrinsic states are highly excited states and thus their properties are almost independent of the nature of the adsorbent;

- no experimental data on the value of the $\tau_p/\tau_{d0}$ ratio are available. That is why all the TPD spectra are simulated both at $\tau_p/\tau_{d0} \approx 1$ and $\tau_p/\tau_{d0} \ll 1$. No essential difference in the features of the spectra is obtained.

The TPD spectra shown in Fig. 1 are obtained with samples of different average size of the Pt clusters. According to Jaeger et al. [5,6] not only the average size is different but also the distribution of the Pt cluster size varies with samples. The size distribution of sample 2 is a narrow one with a single pronounced peak. That is why only the average size is involved in the simulation. The size distribution of samples 4 and 6 is narrow but without any pronounced peak. That is why the details of the distribution enter the simulation. The generalization of Eq. 19 in order to involve the distribution of the Pt cluster size reads:

$$r_d = \sum_{i=1}^{N} P(L_i) r_d(L_i) \qquad (20)$$

where $P(L_i)$ is the statistical weight of the Pt clusters with size $L_i$; $r_d(L_i)$ is given by Eq. 19.

The comparison between simulated and experimental TPD spectra (Figs. 1 and 4) shows a good agreement both in the temperature shift dependence on cluster size and in the broadening of the peaks. The broadening of the peaks with samples 4 and 6 is, however, a superposition between the role of the average size and the role of its distribution. The separation of the two effects is made by extra simulation: a TPD spectrum of sample 6 is simulated separately involving the average size and the size distribution. These simulations are shown in Figs. 5a and 5b. It is easily seen that the broad-ening is due to the lack of any pronounced peak in the size distribution. The difference in the experimentally observed peak area in the TPD spectra in Fig. 5 is explained further.

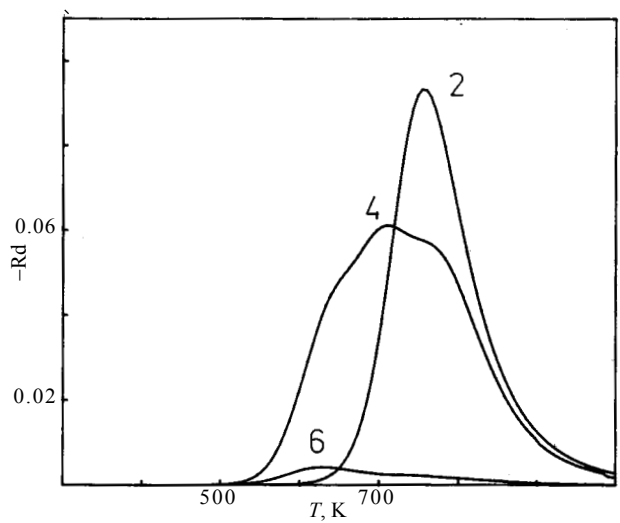

**Fig. 4.** Simulated TPD spectra for samples 2, 4 and 6. Pt cluster sizes and the initial coverage are the same as compared to the corresponding experimental TPD spectra shown in Fig. 1

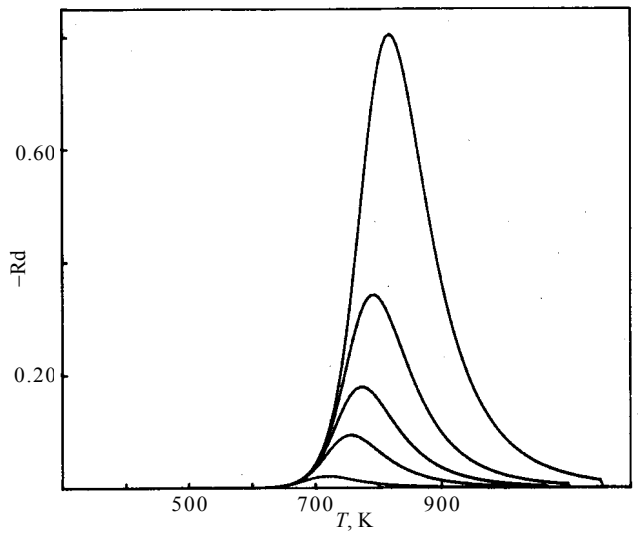

**Fig. 6.** Simulated TPD spectra dependence on the heating rate with sample 6; heating rate (K/min): curve (a) 57.5; curve (b) 23.0; curve (c) 11.5; curve (d) 5.75; curve (e) 1.15

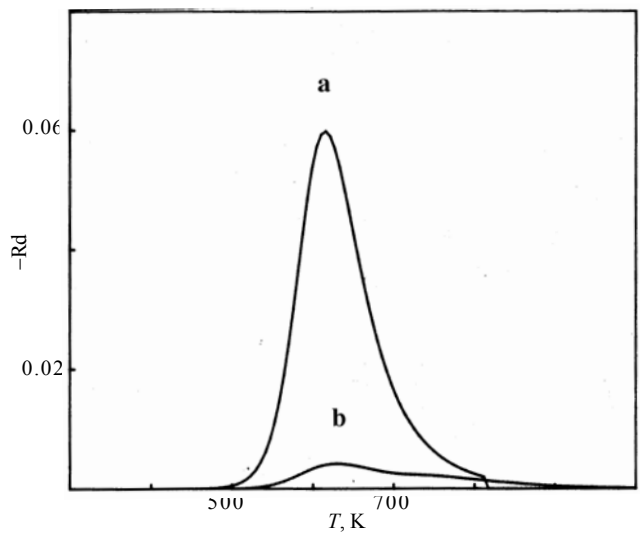

**Fig. 5.** Simulated TPD spectra for sample 6: curve (a) taking into account the average Pt cluster size; curve (b) considering Pt cluster size distribution

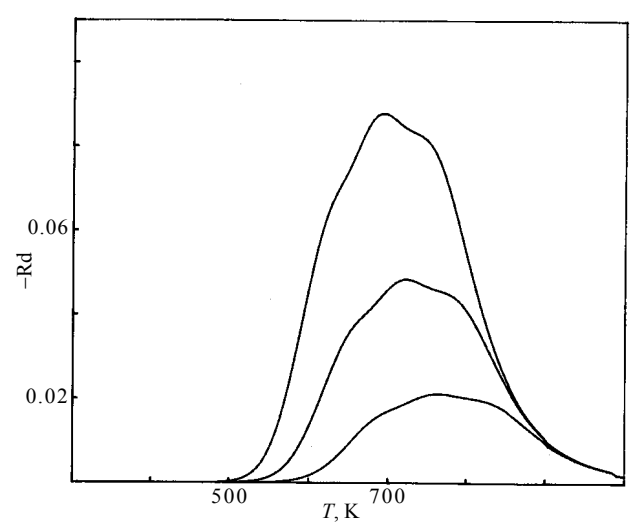

**Fig. 7.** Simulated TPD spectra dependence on the initial coverage with sample 3; initial coverage ($\Theta$ %): curve (a) 34.6; curve (b) 19.8; curve (c) 9.2

The second peculiarity of the experimental TPD spectra is the lack of temperature shift on changing the heating rate. The corresponding simulated TPD spectra are shown in Fig. 6. There is a good agreement between the experimental (Fig. 2) and simulated (Fig. 6) TPD spectra.

The simulated dependence of TPD spectra on the initial coverage is shown in Fig. 7. The simulated TPD spectra demonstrate a slight temperature shift of the peak, contrary to the corresponding experimental TPD spectra shown in Fig. 3. It is supposed that the reason for the discrepancy is in the inaccuracy of acquiring the size distribution of the Pt clusters. According to Jaeger et al. [5,6] the accuracy in the size distribution is 80%.

The final item in this section is the explanation of the difference in the experimentally observed peak area of the spectra for the same initial coverage. This difference appears apparently in Fig. 5 but it is also presented in Fig. 2 and Fig. 6. It is generally accepted that the experimentally observed area in a TPD spectrum is proportional to the initial coverage. This is so, however, only when the desorption rate is a simple exponential function of the inverse temperature. In the present model the desorption rate (Eq. 19) is a complex function of the inverse temperature and it retains long "tails" under the peak. The "tails" are seen in the TPD spectra shown in

Fig. 1. The presence of "tails" is the reason that the observed peak area is not proportional to the initial coverage.

CONCLUSIONS

The major assumption of the model presented is that the adsorption properties on the surface of small metal clusters and at the boundary to the support are different. The effect is pronounced when the adsorption properties on the surface are independent of the particular geometry of the clusters and the main adsorption mechanism is the electronic transfer. The change in the adsorption properties results in the appearance of an extra band in the IR spectra of the adsorbate on small clusters deposited on a non-adsorbing support. An essential property of this band is that its intensity changes on changing the size of the clusters.

An unusual figurative conversion of both types of adsorption sites due to the mobility of chemisorbed and excited molecules causes the formation of a single type of adsorption site, called surface effective site. A distinguished property of the energy of desorption from SES is the explicit dependence on cluster size.

The complicated dependence of the desorption rate on the inverse temperature reveals a lack of proportionality of the observed peak area to the initial coverage.

ACKNOWLEDGEMENT

The authors are grateful to the Bulgarian National Science Fund for financial support under grant X-560.